# ProCAN: Progressive Growing Channel Attentive Non-Local Network for Lung Nodule Classification

Mundher Al-Shabi[a], Kelvin Shak[a], Maxine Tan[a,b]

a) Electrical and Computer Systems Engineering Discipline, School of Engineering, Monash University Malaysia, Bandar Sunway 47500, Malaysia
b) School of Electrical and Computer Engineering, The University of Oklahoma, Norman, OK 73019, USA

Corresponding author: Mundher Al-Shabi (mundher.al-shabi@monash.edu)

## Abstract

Lung cancer classification in screening computed tomography (CT) scans is one of the most crucial tasks for early detection of this disease. Many lives can be saved if we are able to accurately classify malignant/cancerous lung nodules. Consequently, several deep learning based models have been proposed recently to classify lung nodules as malignant or benign. Nevertheless, the large variation in the size and heterogeneous appearance of the nodules makes this task an extremely challenging one. We propose a new Progressive Growing Channel Attentive Non-Local (ProCAN) network for lung nodule classification. The proposed method addresses this challenge from three different aspects. First, we enrich the Non-Local network by adding channel-wise attention capability to it. Second, we apply Curriculum Learning principles, whereby we first train our model on easy examples before hard ones. Third, as the classification task gets harder during the Curriculum learning, our model is progressively grown to increase its capability of handling the task at hand. We examined our proposed method on two different public datasets and compared its performance with state-of-the-art methods in the literature. The results show that the ProCAN model outperforms state-of-the-art methods and achieves an AUC of 98.05% and an accuracy of 95.28% on the LIDC-IDRI dataset. Moreover, we conducted extensive ablation studies to analyze the contribution and effects of each new component of our proposed method.

## 1. Introduction

Cancer is one of the most deadly diseases, whereby one in every six deaths is due to cancer [1]. Most cancer deaths were caused by lung cancer (i.e., 18.4% of all deaths). Furthermore, lung cancer is the most frequently diagnosed cancer (i.e., 11.6% of all cancer cases) [1]. Recent studies [2] show that we can reduce lung cancer mortality by examining patients with Low Dosage Computed Tomography (LDCT) in early stages [3] to capture the cancer heterogeneity in a non-invasive way. However, the task of manually diagnosing patients is extremely time-consuming for the radiologist as every CT scan consists of multiple slices. Furthermore, many lung nodules are very small and benign and malignant nodules sometimes look similar and hard to distinguish from one another.

In recent years, deep learning methods have been shown to significantly outperform the previous hand-crafted feature based approaches due to their ability to extract useful features automatically without relying on hand-crafted features [4–6]. However, their performance for lung nodule classification is still not at the same level as their performance on general classification tasks (e.g., ImageNet).

A main challenge of the lung nodule classification task is that nodule sizes vary greatly, i.e., between 3 to 30 millimeters (mm) in size [7]. This huge size variation combined with the heterogeneity of the nodules makes the task of classifying them [8,9] a very challenging one. A recent study suggested using patches of multiple sizes and implementing multiple feature extractors for each nodule size [10]. However, this solution increases the number of weights and parameters, which is problematic when we train the parameters on small public datasets, such as LIDC-IDRI [7].

Instead of varying the patch sizes [10], we suggested using a Non-Local Network [11] to capture global features that are suitable for middle to big-sized nodules combined with Resnet blocks [12] to capture local features that are suitable for smaller nodules. The combination of Non-Local layers and Resnet blocks produces better results as both global and local features are extracted for the nodule classification task [9]. However, the Non-Local operator does not provide channel-wise attention, which is crucial [13] for feature extraction. Therefore, in this study, we propose a new **C**hannel **A**ttentive **N**on-Local (CAN) network, which generalizes the Non-Local operator to provide channel attention as well as spatial attention.

Another challenge is the trade-off between training a deep network to extract high-level features and the difficulties of training deep networks, such as the vanishing gradient problem. Moreover, the bigger/deeper the network, the higher the required memory and training time. In a recent study, [14] proposed a method called ProGAN to grow the deep network gradually, whereby the network learns low-level features first before discovering high-level features. In this way, the network will need less memory and will converge faster as it is kept small for most training cycles. However, adding new layers during training disturbs the process, at the same time slowing network convergence [14]. Thus, in this paper, we propose to progressively introduce new layers using a novel blending algorithm called the *Bernoulli* Blending algorithm.

We also observed in our previous study [8] that middle-sized nodules (i.e., between 5 to 12 mm) are very challenging to classify. On the other hand, smaller and bigger nodules are easier to classify as benign or malignant [8]. Therefore, to help the deep learning network to learn more effectively, we can gradually and progressively increase the difficulty of the tasks as we grow the network using a method called Curriculum Learning [15]. This method is suitable for lung nodule classification as there are nodules that are easy to classify even by non-complex models, whereas other nodules require complex models to classify correctly [8]. Thus, we first train the network on easier tasks (i.e., easy nodules) using a shallow network; then, we increase the Curriculum's difficulty by training the network on difficult tasks (i.e., hard nodules). Once the

Curriculum gets harder, we increase the network capability by increasing its depth using the new *Bernoulli* Blending algorithm.

To summarize, in this work, we propose a new **Pro**gressively Growing **C**hannel **A**ttentive **N**on-Local (ProCAN) network for lung nodule classification, whereby we add channel attention to the Non-Local Network and propose a new method to increase the depth of the network and the difficulty of the examples gradually. Our main contributions are summarized as follows:

1) The proposed ProCAN model generalizes the Non-Local operator to provide channel attention and spatial attention and integrates them into one block.

2) We enhance the ProGAN gradually growing algorithm by introducing a new *Bernoulli* matrix in the blending algorithm.

3) We use Curriculum Learning with the nodule diameter and radiologist' ratings criteria, whereby we train the network on easy nodules first before the hard ones.

4) The proposed ProCAN model evaluated on the public LIDC-IDRI and LUNGx datasets achieved state-of-the-art performances for both datasets in benign/malignant lung nodule classification.

## 2. Related work

The Non-Local network was first introduced by [11] for video classification and its purpose is to increase the receptive field of the Convolutional Neural Network (CNN) [16] to as large as the input image by considering a weighted sum of all generated features in the whole video. [17] proposed the Transformer, which is based on self-attention operations. The Non-Local operator is closely related to the Transformer [17], which is utilized for language models, but in a two-dimensional (2D) state. Both Transformers and Non-Local networks compute dot-products between the features (see Equations 1-6 in Section 3.1). However, they differ in how these features are represented. For example, in Transformers, the attention is between different words, whereas for Non-Local networks, it is between pixels.

In recent studies, [9] suggested using regular CNN alongside the Non-Local network as the former is suitable for extracting local features only. Therefore, combining Non-Local networks to extract global features combined with CNNs achieves better performance than just regular convolutional models. The Squeeze and Excitation network presented by [13] proposed aggregating the feature maps (Squeeze), followed by a Fully-Connected layer and multiplying the aggregated features to the input (Excitation).

Fu et. al [18] presented their Dual Attention Network that extends the non-local design paradigm for channel attention to spatial attention. The Dual-Attention Network uses two separate and independent attention blocks for channel and spatial attention. Although both Dual Attention and CAN use channel attention, there are three main differences between the two designs: First, Dual Attention uses two separate and independent attention modules, whereas in CAN, the channel attention depends on the spatial attention. Second, CAN uses fewer attention maps to perform the same task as Dual Attention, as shown in Table 1. CAN's parameter-efficient design is important for tasks like lung nodule classification, whereby labeled data is hard to acquire. Third, CAN uses CNNs to solve the permutation equivariant problem in the Non-Local design.

Table 1: Summary and characteristics of related work in spatial and channel attention. N denotes the spatial size; C the number of channels. MM, GAP and FC represent Matrix Multiplications, Global Average Pooling and Fully-Connected, respectively.

| Method | Spatial Attention | Channel Attention | Attention Mechanism | Permutation Equivariant |
|---|---|---|---|---|
| SE [13] | None | $C \times 1$ | GAP with FC | No |
| Non-Local [11] | $N \times N$ | None | MM | Yes |
| Transformer [17] | $N \times N$ | None | MM | No |
| Dual-Attention [18] | $N \times N$ | $C \times C$ | Two Independent MM | Yes |
| Local-Global [9] | $N \times N$ | **None** | MM | Yes |
| 3D Dual-Path [19] | $N \times N$ | **None** | MM | No |
| CAN (ours) | $N \times N$ | $C \times 1$ | Two Dependent MM | No |

With nodule classification being the final stage of CAD schemes for lung cancer, it is not surprising that research is being performed extensively in this field. This might be one of the most important steps in the end-to-end automated schemes for lung cancer prediction in CT scans to classify whether an extracted nodule is malignant or benign.

With the popularity of the non-local paradigm rising, several attempts have been made in the literature incorporating non-local networks. In our recent studies [8,9], we observed that the main difficulty in classifying lung nodules as benign or malignant is due to the large variation in nodule sizes. Therefore, we presented a new method [9] to address this challenge by using non-local mechanisms to extract global features with the help of ResNet blocks to extract local features, and showed that this improved the classification task significantly. In our recent study [8], we proposed another model that revolves around a CNN [16] that utilizes two different dilations and a novel gating sub-network to guide the features between the two dilations.

Other authors such as [19–21] have incorporated models to learn 3D lung nodule characteristics. In a recent study, [19] suggested using a 3D version of the Non-Local network with a Dual-Path network and an ensemble of all the models. [20] performed similar methods of implementing a 3D Dual Path Network to extract feature maps, which then go through the ensembling method of a Gradient Boosting Machine. By considering nine different views of nodule patches, [21] constructed knowledge-based collaborative submodels for each of the views to enhance their classification model.

## 3. Methods

### 3.1. Channel attentive non-local network

In this section, we introduce a new **C**hannel **A**ttentive **N**on-Local (CAN) Network design block as shown in Figure 1 and describe its operation. Firstly, given an input $X \in \mathbb{R}^{C_{in} \times H \times W}$ where $C_{in} \times H \times W$ denote channel, height, and width sizes respectively, we reshaped it to $X \in$

$\mathbb{R}^{C_{in} \times N}$ whereby $N$ is the product of $H$ and $W$. Then we linearly transform $X$ into three feature spaces $Q^T \in \mathbb{R}^{N \times \overline{C}}$, $K \in \mathbb{R}^{\overline{C} \times N}$ and $V \in \mathbb{R}^{C_{in} \times N}$ as shown in equations (1) to (3) below:

$$Q_{j,i} = X_{c,j} M^q_{i,c} \tag{1}$$

$$K_{i,j} = M^k_{i,c} X_{c,j} \tag{2}$$

$$V_{i,j} = M^v_{i,c} X_{c,j} \tag{3}$$

where $M^q \in \mathbb{R}^{\overline{C} \times C_{in}}$, $M^k \in \mathbb{R}^{\overline{C} \times C_{in}}$ and $M^v \in \mathbb{R}^{C_{in} \times C_{in}}$ are learnable parameters and $\overline{C}$ is the number of channels in $Q$ and $K$. The operations in equations (1) to (3) are simple matrix multiplications between matrices $M$ and $X$, which can be implemented efficiently using convolution with a kernel size of one. Then, we create a spatial attention matrix $B \in \mathbb{R}^{N \times N}$ by applying the $Softmax$ function to the product of $Q$ and $K$ as shown in equations (4) and (5) below:

$$S_{i,j} = Q_{i,z} K_{z,j} \tag{4}$$

$$B_{i,j} = Softmax(S_{i,j}) \tag{5}$$

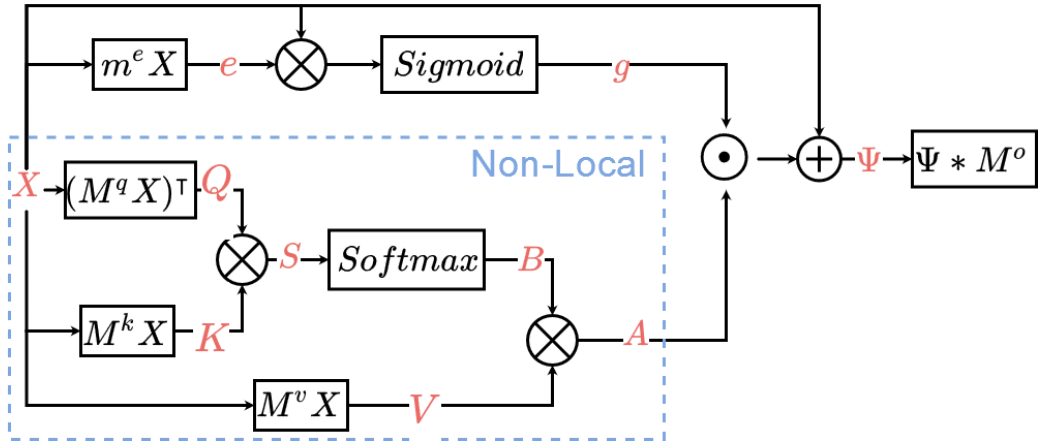

*Figure 1: Block diagram of our new Channel Attentive Non-Local Network (CAN). The block diagram shows the relationship between the standard Non-Local Network (blue dashed box) and our new Channel Attentive operation. Symbols in red are intermediate symbols. ⊗ denotes matrix multiplication, * denotes convolution and ⊙ denotes a multiplication between a vector and a matrix as described in Equation (7).*

For every $N$ spatial features in $B$ we have $N$ attention values whereby $B_{i,j}$ attends to the ith spatial feature for the jth output location. Finally, we multiplied the attention matrix $B$ with the feature space $V$ in equation (6) as follows:

$$A_{c,j} = V_{c,i} B_{i,j} \tag{6}$$

where $A = (a_{c1}, a_{c2}, \dots, a_{C_{in}})$ and $V = (v_{c1}, v_{c2}, \dots, v_{C_{in}})$. Every value in the $a_c \in \mathbb{R}^N$ has been calculated from all values in $v_c \in \mathbb{R}^N$, while $B$ represents the spatial attention of each value. Note that the same values in $B$ are applied over all channels in V, namely the attention mechanism does not change from one channel to another. Hence, the attention mechanism here is called spatial attention as the attention varies across different positions (or pixels) in the image, but the same attention is applied across all channels. Compared with the conventional convolutional neural network (CNN) [16], whereby the receptive field is limited to the kernel size (typically, 3×3, 5×5, 7×7, etc.), the Non-Local Network's receptive field has the same dimensions as the input $X$. Thus, we applied the concept of the Non-Local Network in our CAN, which captures long-range dependencies and computes the attention (or response) at a pixel as a weighted sum of the features at all pixels.

The Non-Local Network only considers spatial (or position) based attention to the image. Thus, to extend the attention mechanism to the channels of the image, we introduced and applied a novel Channel Attentive operation on A with residual connection coming from $X$:

$$\Psi_{c,n} = A_{c,n} g_c + X_{c,n} \quad (7)$$

In contrast with B, the spatial attention matrix, $g \in \mathbb{R}^{C_{in}}$ is the attention vector for channels where $g_c$ is the $c^{th}$ attention value for the $c^{th}$ channel in A. Similar to the spatial attention, we used the Non-Local paradigm to generate $g$. To do so, we first created a feature space $e \in \mathbb{R}^N$ by multiplying $X$ with a learnable vector $m^e \in \mathbb{R}^N$:

$$e_j = m_c^e X_{c,j} \quad (8)$$

We then applied the $Sigmoid$ function to the product of $X$ and $e$, to limit the $g$ values to range between zero and one as follows:

$$g_c = Sigmoid(X_{c,j} e_j) \quad (9)$$

The attention mechanism here is a novel channel attention mechanism that we propose in this work. $g_c$ is the attention mechanism for different channels and is an extension of the Non-Local spatial attention paradigm applied to channel attention.

The channel attention, $g_c$ is multiplied with the spatial attention matrix, $A$. Then, at the end of the CAN layer, we add this result to the input, $X$ to obtain matrix, $\Psi$. The $\Psi$ matrix is reshaped from $\mathbb{R}^{C_{in} \times N}$ to $\mathbb{R}^{C_{in} \times H \times W}$ and passed through a 3x3 convolution with a $relu$ activation function:

$$O = relu(\Psi * M^o) \quad (10)$$

where $M^o \in \mathbb{R}^{C_{out} \times C_{in} \times 3 \times 3}$ and $C_{out}$ is the number of output channels. Applying the two-dimensional (2D) 3x3 convolution allows us to control the number of the output channels and the spatial dimension. Moreover, adding a local operator like the 2D convolution to the Non-Local design block increases the accuracy of the CAN network, which has been shown and experimented in multiple works [9,22]. By incorporating both local and global (i.e., non-local) layers to the network architecture, both local and global features can be extracted in an image, which have been shown to be very useful for lung nodule classification in our previous study [9]. This is mainly because lung nodule classification is a very challenging task as lung nodules have diverse shapes and sizes [23]. Global feature extraction through the Non-Local Network is required to extract features that can describe the nodule shape and size, whereas local features are important to pay attention to details including the nodule density and texture.

## 3.2. Curriculum learning method for hard and easy-to-classify nodules

Inspired by humans and animals, in curriculum learning, the training examples are ordered in a meaningful way [15]. Remarkably, the examples are ordered from easy to hard where the model is trained first on easy examples before the hard ones. Such a strategy helps the optimizer to find a better local minima and converge faster [15].

We propose to train the model on the easy examples first, and we stop the training if the accuracy of the next epoch is lower than the accuracy of the last three epochs:

$$acc_{j+1} < min(acc_j, acc_{j-1}, acc_{j-2}) \tag{11}$$

where $acc_j$ is the validation accuracy for the $j^{th}$ epoch. The validation set is independent of the training and testing sets, whereby we sampled 10% of the training set for the validation set.

For our lung nodule classification task, there are two criteria we can use to classify an example as either easy or hard. The first criterion is based on the median score from the four radiologists (see Section 4.1.1), whereby on a scale of 1 to 5, 1 indicates that a nodule is benign and 5 that it is malignant. The easy examples are the ones that are clearly benign (rating 1) or clearly malignant (rating 5) [24]:

$$\text{difficulty}_i = \begin{cases} easy, & r_i = 1 \text{ or } r_i = 5 \\ hard, & otherwise \end{cases} \tag{12}$$

where $r_i$ is the median rating of the $i^{th}$ training example. This criterion assumes that what is not clearly benign or malignant from the radiologists' perspective would be hard for Artificial Intelligence (AI) to classify. However, this assumption can be challenged by many examples in the real-world, whereby AI can solve many tasks easily that have been recognized as hard by domain experts and vice versa [25,26].

The second criterion is based on the diameter of the nodule. Usually, small nodules (i.e., <5mm) are benign, whereas big nodules (i.e., >12mm) are malignant [8]. The classifier can be easily trained to distinguish these nodules (in fact, by applying a simple thresholding operator, one can likely obtain a reasonable nodule classification score [8]). However, the nodule sizes between 5 to 12 mm represent the mid-range nodules that are "difficult" to classify and there is no straightforward way to classify these nodules [8]:

$$\text{difficulty}_i = \begin{cases} hard, & 5 \text{ mm} \leq d_i \leq 12 \text{ mm} \\ easy, & otherwise \end{cases} \tag{13}$$

where $d_i$ is the $i^{th}$ diameter of the training example. In contrast with the first criterion, we applied this criterion from the perspective of AI instead of the radiologists' perspective. Namely, we observed in our previous experiments [8] that AI performs really well in classifying nodules <5 mm (as benign) and >12 mm (as malignant); however, its performance is poor in classifying the mid-range nodules between 5 to 12 mm.

*3.3. Progressive growing method with a two-dimensional Bernoulli matrix*

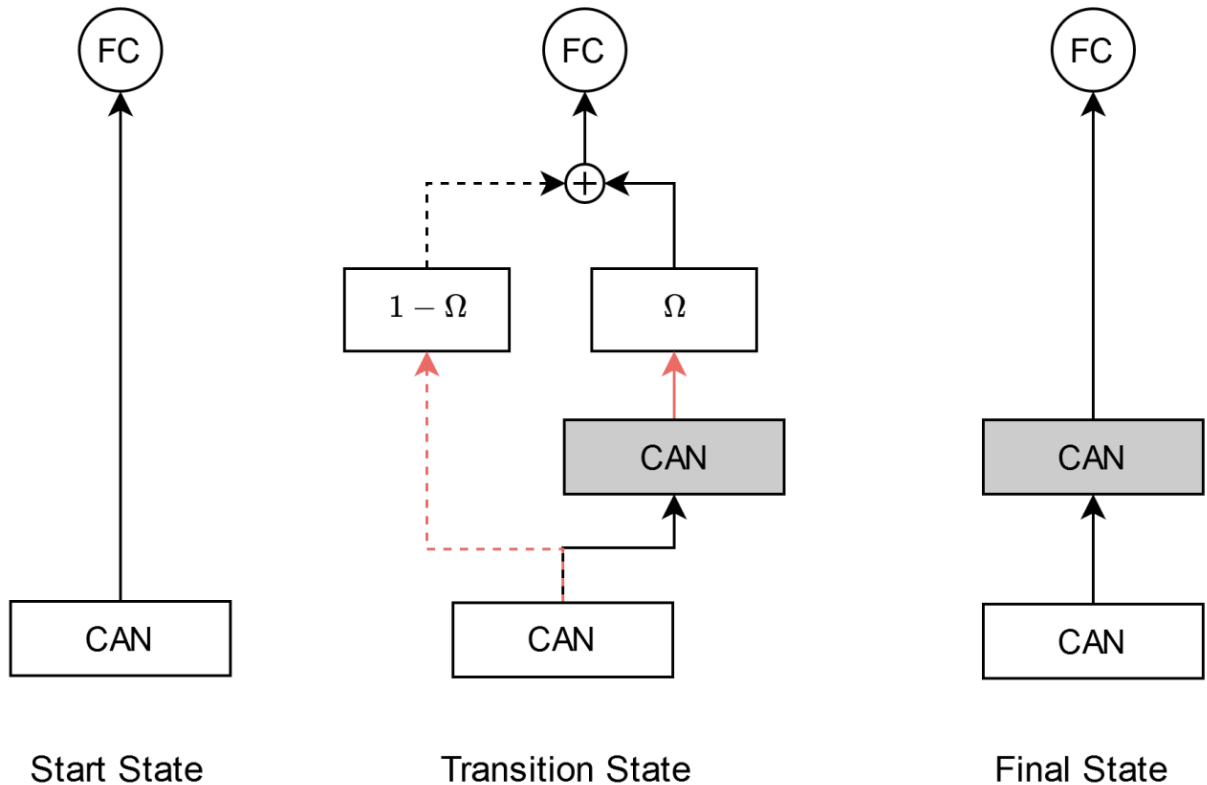

Figure 2: Progressive Growing States from start state, transition state to final state. The dashed connection is temporary and used during the transition state. Once we reach the final state, we may remove the temporary connection. The Channel Attentive Non-Local (CAN) block with gray color is the new CAN block, and we refer to it in the text (Section 3.3) as $\boldsymbol{\theta}$ while the CAN block with white background is a part of the base feature extractor, which we refer to in the text as $\boldsymbol{f_0}$. The red arrows denote multiplications while the black arrows are just simple connectors.

In this section, we propose a new Progressive Growing method to help increase or progressively grow our network architecture in steps. The performance of a deep learning network is limited by its size/architecture [12]; therefore, to ensure that the deep learning network performs optimally, we need to gradually increase its size and progressively grow it in a "safe" environment. As we increase the difficulty of the task using Curriculum Learning (in Section 3.2), the network performance reaches its limit, and the accuracy cannot be improved further on the current network architecture. Therefore, we propose to add new CAN blocks on the fly during the training procedure.

Our network consists of two main functions: base feature extractors, $\boldsymbol{f_0}$ (CANs) and a classifier $\boldsymbol{u}$ (i.e., a fully-connected layer). The feature extractor $\boldsymbol{f}$ consists of one or more CAN blocks, $\boldsymbol{\theta}$ as follows:

$$\boldsymbol{f_0} = \boldsymbol{\theta}_l(\boldsymbol{\theta}_{l-1}(\ldots, \boldsymbol{\theta}_1(\boldsymbol{x}))) \qquad (14)$$

where $l$ is the number of CAN blocks, and $\boldsymbol{x}$ is the input image. The straightforward implementation would be connecting the feature extractor with a fully-connected layer directly as $u(f(x))$. To expand the base feature extractor $f$, we can simply attach a new CAN block $\theta$ to the end of $f$.

$$f_i(x) = \theta_i(f_0(x)) \tag{15}$$

where $\theta_i$ is the $i^{th}$ CAN block that is being added during the training to $f_i(x)$. Nevertheless, the process of adding a new block during the training will disturb the training procedure, and the optimizer may take longer to converge or it may be stuck at bad local minima [14]. Therefore, for the Progressively Growing GAN (ProGAN) architecture [14], gradual growing was proposed by introducing a scalar value $0 \leq p_i \leq 1$ to control the flow of the information from the base features $f_0$ and the new CAN block $\theta_i$ as follows:

$$f_i(x) = p_i \theta_i(f_0(x)) + (1 - p_i) f_0(x) \tag{16}$$

The gradual growing process passes through three states: (1) start state, (2) transition state, and (3) final state. In the start state, the $p_i$ value is set to $p_i = 0$ to block all the features from the new block $\theta_i$; namely, in the start state, it is as though there is no new block added to the network architecture. Then, in the transition state, the $p_i$ values are set to $p_i = \{0.25, 0.5, 0.75\}$ in that order. For example, in the first epoch, $p_i = 0.25$; then, in the second epoch $p_i = 0.5$, and so on. This allows the new features that come from $\theta_i$ to slowly blend with the old features that come from $f$. Namely, the network architecture is increased in a "safe" environment to allow the network to gradually adapt to prescribed changes in the architecture. In the final state, the $p_i$ value is set to $p_i = 1$ to allow all features from $\theta_i$ to pass through, as depicted in Figure 2.

In the ProGAN paper [14], it was shown that gradually introducing $\theta_i$ improves the network's performance. However, the main shortcoming of this approach is that multiplying with the $p_i$ values modifies/destroys every single value coming from $\theta_i$. As we modify $\theta_i$ and $f_0(x)$ in every epoch, the following fully-connected layer, $u$ has to adapt to this shift of the feature values in every single epoch. For example, when $p_i = 0.5$, this means every value coming from $\theta_i$ is divided by two. The multiplication with $p_i$ modifies the feature maps which leads to losing the information gained by $\theta_i$ or $f_0(x)$. To overcome this problem and to avoid modifying the feature values, we propose to use a 2D **_Bernoulli_** matrix with a probability of $p_i$ instead of just multiplying with the scalar values, $p_i$ as follows:

$$\Omega_i \sim Bernoulli(p_i) \tag{17}$$

where $\Omega_i \in \mathbb{R}^{H \times W}$ is a 2D matrix that has the same spatial size as the output features from $f_0(x) \in \mathbb{R}^{C_{in} \times H \times W}$.

Every value in the **_Bernoulli_** matrix $\Omega_i = \{\omega_{1,1}, \cdots, \omega_{H,W}\}$ is either zero or one and when we multiply it with $\theta_i$, the **_Bernoulli_** matrix either allows the feature to pass through completely (i.e., $\omega = 1$) or blocks it completely (i.e., $\omega = 0$), without modifying the feature values in any way. In this way, applying the **_Bernoulli_** matrix does not modify the output values unlike multiplying with the $p_i$ values in the ProGAN paper [14]:

$$f_i(x) = \Omega_i \theta_i(f_0(x)) + (1 - \Omega_i)f_0(x) \tag{18}$$

Finally, we connect $f_i$ to the last layer, which is a fully-connected layer, $u$ to obtain the final output of our new ProCAN network:

$$\tilde{y}_i(x) = u(f_i(x)) \tag{19}$$

## 3.4. Full algorithm and neural network architecture

The pseudocode of our ProCAN method is given in Algorithm 1. We first train the network on easy training data, followed by hard training data in the Curriculum Learning framework (i.e., Section 3.2) by minimizing the Binary Cross Entropy (BCE) loss function. We then progressively grow the network by adding new CAN blocks using our new Progressive Growing approach (i.e., Section 3.3). Finally, the number of CAN blocks, $T$ that gives the best result and its corresponding output, $\tilde{y}_T$ are returned at the end of Algorithm 1.

---

*Algorithm 1: ProCAN training procedure*

**Requirements:** $\{x_{easy}, y_{easy}\}$: easy training data, $\{x, y\}$: easy and hard training data, $f_0$: base feature extractor, $u$: fully connected layer, $T$: number of extended CAN blocks

**minimize** $loss(u(f_0(x_{easy})), y_{easy})$ until Eq. 11 satisfied
**minimize** $loss(u(f_0(x)), y)$ until Eq. 11 satisfied
**for** $i \in \{1, \ldots, T\}$ **do**
  $\theta_i = $ **new** $CAN\ block$
  **for** $p_i \in \{0.25, 0.5, 0.75, 1\}$ **do**
    $\Omega_i \sim Bernoulli(p_i)$
    $f_i(x) = \Omega_i \theta_i(f_{i-1}(x)) + (1 - \Omega_i)f_{i-1}(x)$
    $\tilde{y}_i(x) = u(f_i(x))$
    **minimize** $loss(\tilde{y}_i(x), y)$ for one epoch
  **end for**
  **remove** unnecessary temporary connectors from $\tilde{y}_i$
  **minimize** $loss(\tilde{y}_i(x), y)$ until Eq. 11 satisfied
**end for**
**minimize** $loss(\tilde{y}_T(x), y)$ for remaining epochs
**Return:** $\tilde{y}_T$

---

The final network structure of our ProCAN method consists of seven blocks of CANs followed by a Global Average Pooling (GAP) layer and a Fully-Connected layer, as tabulated in Table 2 (see Ablation study results in Section 5.3 for a full analysis of how seven CAN blocks were obtained for the final network architecture). The first four layers are a part of the base feature extractors (see the first part of the Ablation study results in Section 5.3 for a full analysis of how 4 base feature extractors was obtained), whereas the other three are added during the training (see the second part of the Ablation study results in Section 5.3 for a full analysis of how 3 extended feature extractors was obtained). The input to the first base feature extractor is a 3D volume of size 32×32×32 mm³ at the nodule center. The Stride in Table 2 refers to the stride size of the last convolution operation in the CAN block (i.e., Eq. 10). We also set the size of the intermediate channels, $\bar{C}$ (in matrices $M^q$ and $M^k$ in Eqs. 1 and 2, respectively) of the CAN blocks to one (see the second part of the Ablation study results in Section 5.2 for a full analysis of how the optimal number of intermediate channels, $\bar{C} = 1$ was obtained).

Table 2: Network Architecture of ProCAN

| Layer | Input $C_{in} \times H \times W$ | Output $C_{out} \times H \times W$ | Stride |
|---|---|---|---|
| Base CAN 1 | 32×32×32 | 32×32×32 | 1 |
| Base CAN 2 | 32×32×32 | 64×16×16 | 2 |
| Base CAN 3 | 64×16×16 | 128×8×8 | 2 |
| Base CAN 4 | 128×8×8 | 256×8×8 | 1 |
| Extended CAN 1 | 256×8×8 | 256×8×8 | 1 |
| Extended CAN 2 | 256×8×8 | 256×8×8 | 1 |
| Extended CAN 3 | 256×8×8 | 256×8×8 | 1 |
| GAP | 256×8×8 | 256×1×1 | ------ |
| Fully-Connected | 256×1×1 | 1 | ------ |

We also placed a batch normalization layer [27] at the end of every CAN block to normalize the feature maps. Moreover, we applied dropout regularization [28] with a probability of 0.5 before the Fully-Connected layer to prevent overfitting and to improve network generalizability.

## 3.5. Preprocessing

Our first preprocessing step was to normalize the CT scans using trilinear interpolation. This results in isotropic resolution in all three (*x*, *y*, and *z*) dimensions. Then we cropped a volume of size 32×32×32 mm$^3$ around the nodule, which is enough to fit the largest nodule (see Section 4.1 and the largest diameter of nodules in the LIDC-IDRI dataset is 30 mm). Third, we clamped the Hounsfield unit (HU) values of the scans that were less than -1000 or larger than 400, which is a common practice employed in the literature to filter out the air and bone regions [8,10], as follows:

$$x = min(max(x, -1000), 400) \qquad (20)$$

Finally, we normalized the nodules to have zero mean and unit variance according to standard practice in the deep learning literature.

## 3.6. Data augmentation

A method that is frequently used in the literature to avoid or prevent overfitting of a deep learning model is augmenting the training data. In this study, we augmented the nodules in our dataset by rotating each nodule around the three (*x*, *y*, and *z*) axes. For each axis (*x*, *y*, and *z*), we rotated each nodule in seven directions (i.e., 0º, 45º, 90º, 135º, 180º, 225º, 270º). This means that every training example was augmented 21 times (i.e., 3×7).

Although data augmentation regulates the model and prevents it from overfitting, it also introduces a gap in the data distribution between the training and test data [29]. Therefore, we experimented with refining the model at the end of the training procedure by training the model on non-augmented data; that is, we trained the model on augmented data first, and then refined the trained model without any data augmentation. Unless explicitly mentioned in the manuscript, all proposed methods were trained on 9 folds with data augmentation and tested on the remaining one fold without data augmentation.

## 4. Experiments and results

### 4.1. Datasets

#### 4.1.1. LIDC-IDRI dataset

The Lung Image Database Consortium and Image Database Resource Initiative (LIDC-IDRI) database [7] is the largest publicly-available lung nodule dataset. It contains 1,018 computed tomography (CT) scans from 1,010 patients altogether collated from seven academic centers across the United States (US). The slice thicknesses of the CT scans range from 0.45 to 5.0 mm, as the scans were collated from different institutions and devices.

Table 3: Summary of distribution of CT scans/patients in the two publicly-available LIDC-IDRI [7] and LUNGx [30] datasets that were used in this study. LIDC-IDRI and LUNGx are highly-popular and LIDC-IDRI is the biggest publicly-available dataset for lung cancer classification.

| Public dataset | Number of malignant cases | Number of benign cases |
|---|---|---|
| LIDC-IDRI | 406 | 442 |
| LUNGx | 36 | 37 |

Each CT scan was annotated by four experienced thoracic radiologists altogether. In this study, we follow a rigorous approach as [8,23], whereby we only used the nodules that were annotated by at least three out of four radiologists to ensure that the majority of radiologists agreed on the presence of a nodule. The four radiologists also rated each nodule on a scale of one to five, whereby one indicates that a nodule is clearly benign and five, clearly malignant. Therefore, to aggregate all of the malignancy ratings by all four radiologists, similar to previous studies [8,9,32], we took the median of the ratings of all four radiologists as the ground truth in our experiments. We excluded nodules with a median rating of 3 as we could not assign them to either benign or malignant groups. In this way, we obtained 848 nodules altogether, of which 442 were benign and 406 malignant as shown in Table 3.

#### 4.1.2. LUNGx dataset

The LUNGx Challenge dataset associated with the 2015 SPIE Medical Imaging Symposium challenge [30] consists of 70 CT scans – 10 for calibration and 60 for testing. The 60 test scans contain 73 nodules, of which 37 are benign, and 36 are malignant as shown in Table 3. Considering there is no training set associated with the LUNGx dataset, we used the LIDC-IDRI dataset for training purposes, as suggested in the challenge guidelines [30]. Similar to other studies [30], we only examined the performance of our method on the testing set.

### 4.2. Experimental setup

Similar to our previous studies [8,9] and other studies [10,32,33] conducted in the literature, we evaluated our proposed ProCAN model on the LIDC-IDRI dataset using a 10-fold cross-validation method, in which 9 folds were used for training and one fold for testing. Moreover, we sampled 10% of the training data as a validation dataset.

For the second dataset, namely LUNGx, no training dataset was provided. Therefore, similar to previous studies that utilized this dataset [30,33], we directly tested our trained ProCAN model on the 73 nodules in the testing dataset. To validate our model on the LUNGx dataset, we used the same validation method utilized by the organizers of the LUNGx Challenge [30], namely bootstrapping with 1000 iterations.

Our ProCAN network implementation was based on PyTorch 1.4 [34]. We ran our experiments using a NVIDIA RTX 2080Ti GPU. The following parameters apply for our experimental design: We applied a weight decay of 0.0001 for the fully-connected layer. We also set the weight decay and dropout values to 0.5 to regularize the fully-connected layer and prevent it from overfitting. For the learning rate, we first initialized it to 0.001 and then to 0.0001 at the 20$^{th}$ epoch. All the models were optimized using the Adam optimizer [35] for 60 epochs with a batch size of 256. At the 51$^{st}$ epoch, the model was refined by training it without augmented data.

### 4.3. Evaluation criteria

The performance of the lung nodule classification scheme was assessed using the standard evaluation criteria used in the literature, namely accuracy, sensitivity, precision, and AUC. These performance metrics were computed using equations (21) to (25) below:

$$\text{Accuracy} = \frac{TP + TN}{TP + TN + FN + FP} \quad (21)$$

$$\text{Sensitivity} = \frac{TP}{TP + FN} \quad (22)$$

$$\text{Precision} = \frac{TP}{TP + FP} \quad (23)$$

$$F1 - \text{Score} = \frac{2TP}{2TP + FP + FN} \quad (24)$$

$$\text{AUC} = \int_0^1 t_{pr}(f_{pr}) df_{pr} = P(X^+ > X^-) \quad (25)$$

where $TP$, $TN$, $FN$, and $FP$ represent the number of true positive, true negative, false negative, and false positive nodules, respectively. In a ROC curve, the false positive rate (i.e., $f_{pr}$) axis ranges from 0 to 1 and the true positive rate (i.e., $t_{pr}$) is a function of the $f_{pr}$; $X^+$ and $X^-$ represent the confidence/probability scores for a positive and negative sample, respectively.

### 4.4. Results

In this section, we compare our proposed model with state-of-the-art models in the literature on both popular datasets of LIDC-IDRI and LUNGx. We first compare our method with other methods on the LIDC-IDRI database, then we perform the same analysis on the LUNGx database.

Table 4: Comparison of our ProCAN and Ensemble ProCAN methods with other state-of-the-art methods in the literature on the LIDC-IDRI database.

| Model | Ensemble? | AUC | Accuracy | Precision | Sensitivity | F1-Score |
| --- | --- | --- | --- | --- | --- | --- |
| Multi-Crop [32] | No | 93.0 | 87.14 | --- | 77.0 | --- |
| HSCNN [36] | No | 85.6±2.6 | 84.2±2.5 | --- | 70.5±4.5 | --- |
| Local-Global [9] | No | 95.62±0.02 | 88.46±0.04 | 87.38±0.07 | 88.66±0.06 | 88.37±0.04 |
| Gated-Dilated [8] | No | 95.14±0.03 | 92.57±0.03 | 91.85±0.05 | 92.21±0.04 | 92.60±0.03 |
| Swarm [37] | No | --- | 93.71 | 93.53 | 92.96 | --- |
| ProCAN (ours) | No | **97.13±0.02** | **94.11±0.03** | **94.54±0.04** | **93.12±0.05** | **93.81±0.03** |
| 3D DPN(Ensemble)[19] | Yes | --- | 90.24 | --- | 92.04 | 90.45 |
| MV-KBC [21] | Yes | 95.70±0.24 | 91.60±0.15 | 87.75±0.24 | 86.52±0.25 | 87.13±0.16 |
| MSCS-DeepLN [10] | Yes | 94.00±0.25 | 92.65±0.26 | 90.39±0.93 | 85.58±0.94 | 87.91±0.43 |
| MK-SSAC [33] | Yes | 95.81±0.19 | 92.53±0.05 | --- | 84.94±0.17 | --- |
| Ensemble ProCAN (ours) | Yes | **98.05±0.02** | **95.28±0.02** | **95.75±0.04** | **94.33±0.04** | **95.04±0.02** |

In the first 6 rows of Table 4, we analyze the performance of our ProCAN model against other methods on the LIDC-IDRI database. Moreover, we designed an ensemble version of ProCAN (i.e., Ensemble ProCAN) to compare with recent ensemble models in the literature [10,19,21,33]. We constructed our ensemble model by taking a simple average sum of three different variations of ProCAN: ProCAN-6, ProCAN-7, and ProCAN-8, where 6, 7, and 8 denote the number of the blocks used in the ProCAN network architecture. The individual performances results of ProCAN-6, ProCAN-7, and ProCAN-8 are given in Section 5.3.

From Table 4, we observe that both our ProCAN and Ensemble ProCAN models outperform the state-of-the-art models in the literature on all evaluation criteria. Also, our Ensemble ProCAN model outperforms ProCAN on all the evaluation criteria, which is consistent with the results in the literature that show that ensemble methods generally outperform non-ensemble ones [19,21]. Furthermore, our non-ensemble ProCAN model marginally outperforms all the non-ensemble and ensemble methods in Table 4, excluding Ensemble ProCAN, whereas Ensemble ProCAN considerably outperforms all other methods in Table 4.

We also compare the performance of our ProCAN and Ensemble ProCAN models with other state-of-the-art models on the LUNGx database. With the LUNGx database, no training dataset is provided; therefore, similar with other methods that use this database, we only evaluate our proposed model on a provided testing dataset. Thus, the evaluation of our proposed model on this dataset measures the robustness of the model on changes in the distribution of the applied dataset, namely it evaluates the generalizability of our model on completely independent/unseen datasets. The results and comparisons with other state-of-the art methods [21,30] are tabulated in Table 5.

Table 5: Comparison of our ProCAN method with other state-of-the-art and best-performing methods on the LUNGx Challenge dataset. We have incorporated these results from the relevant published papers [21,30]. .

| Method | AUC |
| --- | --- |
| Voxel-intensity-based segmentation + SVM | 50±6.8 |
| Region growing + WEKA | 50±5.6 |
| Rules based on histogram-equalized pixel frequencies | 54±6.7 |
| Bidirectional region growing + Uses tumor perfusion surrogate | 54±6.6 |
| Region growing + WEKA | 55±6.7 |
| Graph-cut-based surface detection + Random forest | 56±5.4 |
| Manual initialization, gray-level thresholding, morphological operations + SVM | 59±6.6 |
| Convolutional neural network | 59±5.3 |
| GrowCut region growing with automated initial label points + SVM | 61±5.4 |
| Radiologist-provided nodule semantic ratings + Discriminant function | 66±6.3 |
| Semiautomated thresholding + Support vector regressor | 68±6.2 |
| MV-KBC [21] | 76.85±0.17 |
| ProCAN (ours) | **77.2±5.1** |
| MK-SSAC [33] | 78.83±0.75 |
| Ensemble ProCAN (ours) | **80.1±4.6** |

We observe in Table 5, that our ProCAN and Ensemble ProCAN models achieve the highest AUC values compared to standard machine learning and state-of-the-art methods [21,30]. In particular, Ensemble ProCAN achieves AUC = 80.1±4.6, which to the best of our knowledge is a result that surpasses all other reported methods in the literature on this challenging dataset to date. These results also demonstrate that ProCAN and Ensemble ProCAN are generalizable to unseen datasets and their performance is robust to variability in the distribution of the dataset at hand.

## 5. Discussion

In this section, we analyze the efficacy of our proposed model, namely each proposed component of our ProCAN network architecture. We perform extensive studies and compare the performance of our methods with other established state-of-the-art methods in the literature. We also conduct extensive ablation studies to examine each applied step and component described in the Methods section of this paper.

### 5.1. Analysis of different data augmentation methods

We analyze our proposed algorithm under different data augmentation settings in this section. First, we examine the performance of our method without any data augmentation to verify that applying data augmentation improves the results. Next, we analyze our method's performance with different rotation angles (i.e., seven and four rotation angles). Finally, we examine the application of the refinement technique described by [29] and explained in Section 3.6. We tabulated all the obtained results in Table 6.

Applying the refinement technique [29] was beneficial for a data augmentation with both four and seven angles. Especially for data augmentation with seven angles, the refinement method improved the results of all analyzed performance metrics; for data augmentation with four angles, improvements were observed for all performance metrics except the sensitivity, which declined marginally by 0.49%. Overall, the best result was obtained for data augmentation with seven angles with refinement.

Table 6: Results of different data augmentation methods. All these models were trained for 60 epochs. 'Refine' denotes training the model with augmented data for 50 epochs, then training for an additional 10 epochs without data augmentation. '7 Angles + Test Aug' denotes applying 7 augmentation angles during training and testing. 'None' denotes that no augmentation nor refinement was applied.

| Augmentation Method | AUC | Accuracy | Precision | Sensitivity | F1-Score |
|---|---|---|---|---|---|
| 7 Angles + Refine | **97.13** | **94.11** | **94.54** | **93.12** | **93.81** |
| 7 Angles | 96.54 | 93.87 | 94.25 | 92.86 | 93.55 |
| 4 Angles + Refine | 96.37 | 92.92 | 92.61 | 92.61 | 92.61 |
| 4 Angles | 96.19 | 92.81 | 91.97 | 93.1 | 92.53 |
| 7 Angles + Test Aug | 97.10 | 93.99 | 94.26 | 93.10 | 93.68 |
| 7 Angles + Test Aug + Refine | 96.92 | 93.87 | 94.25 | 92.86 | 93.55 |
| None | 93.31 | 88.56 | 89.31 | 86.45 | 87.86 |

We observe in Table 6 that the performance drops significantly when we do not apply any data augmentation. The results for seven and four rotation angles are quite close, whereby the augmentation with seven angles improves the accuracy by ~1% and improves all the other performance metrics except the sensitivity result, which declines marginally by 0.14%.

Furthermore, we tested the augmentation at test time as a replacement for refinement. First, we conducted an experiment, namely "7 Angles + Test Aug", whereby we augmented the test data 27 times (7 angles × 3 views) and averaged the results. As shown in Table 6, it performs slightly worse than applying augmentation and refinement ("7 Angles + Refine"). Besides the marginal gain in AUC and Accuracy when using refinement rather than augmenting the test data, refinement is 27 times faster as we just pass the image once during test time. We also combined all methods "7 Angles + Test Aug + Refine", but the result is even worse than using refinement or augmenting the test data, since in the last 10 epochs of the refinement, we tune the model on one angle and one view, but test it on different angles and views.

## 5.2. Design of CAN block

In this section, we analyze the importance of channel attention and our CAN design block in an extensive ablation study. First, we remove the channel attention from the CAN block to examine the effectiveness of our new channel attention mechanism described in Section 3.1. In doing so, we are left with the Non-Local network (i.e., the blue dashed box in Figure 1) and the last convolutional layer (i.e., equation 10) in our network architecture. This constitutes the first result obtained in the first row of Table 7 (i.e., the "Non-Local" model).

To examine the efficacy of our new CAN block even further, we also examine two different state-of-the-art and popular channel attention mechanisms: (1) Squeeze and Excitation (SE) Networks [13] and (2) Dual Attention Networks [18]. The SE Network weights is a highly popular network that weights each channel to a single numeric value using average pooling. Dual Attention Networks on the other hand, use the Non-Local paradigm for spatial and channel attention, but with two separate blocks.

We ran the experiments and compared all three modifications of channel attention with our CAN block and tabulated the results in Table 7. From the results, we observe that adding the channel attention in the form of the SE block combined with the Non-Local network improves the performance across all performance metrics, as expected. The Dual Attention Network performs better than the Non-Local Network individually, but its performance is poorer than Non-Local combined with SE. This is due to the redundancy in the Dual Attention Network design block, whereby there are two independent skip connections in the block (see Fig. 2 of the [18] reference). Another shortcoming of the Dual Attention Network is the size of the intermediate channels $\bar{C} = C_{in}$, which not only slows the training but also increases the number of learnable parameters (please refer to the analysis of the results of Table 8 below where we performed the ablation study of the number of learnable parameters). We observe in Table 7 that apart from the sensitivity result (which is marginally below Non-Local plus SE by 0.48%), our new CAN block outperforms all the other methods across all performance metrics.

*Table 7: Performance comparisons of our novel CAN block design with other popular or state-of-the-art methods in the literature*

| Model | AUC | Accuracy | Precision | Sensitivity | F1-Score |
|---|---|---|---|---|---|
| Non-Local | 96.4 | 92.81 | 93.02 | 91.87 | 92.44 |
| Non-Local + SE | 96.56 | 93.75 | 93.37 | **93.6** | 93.48 |
| Dual Attention | 96.48 | 93.16 | 93.07 | 92.61 | 92.84 |
| CAN | **97.13** | **94.11** | **94.54** | 93.12 | **93.81** |

We also studied the effects of varying the number of intermediate channels, $\bar{C}$ described in Section 3.1 on the overall results. $\bar{C}$ is the number of channels in the $Q$ and $K$ matrices described in equations (1) and (2), respectively in Section 3.1. We examined three different $\bar{C}$ values altogether; first, $\bar{C} = C_{in}$ as prescribed by [18], which simply means that the number of intermediate channels should be equal to the number of input channels. Second, [38] proposed to set $\bar{C} = C_{in}/8$, which means that $\bar{C}$ is eight times smaller than $C_{in}$, thus reducing computational requirements and the overall number of learnable parameters in the process. The third value of $\bar{C}$ that we tried was just $\bar{C} = 1$, which is the minimum value of $\bar{C}$ that can be used. We tabulated the obtained results in Table 8.

*Table 8: Effects of varying the number of intermediate channels*

| Number of intermediate channels, $\bar{C}$ | AUC | Accuracy | Precision | Sensitivity | F1-Score |
|---|---|---|---|---|---|
| 1 | **97.13** | **94.11** | **94.54** | 93.12 | **93.81** |
| $C_{in}/8$ | 96.84 | 93.87 | 93.38 | **93.84** | 93.61 |
| $C_{in}$ | 96.85 | 92.92 | 93.03 | 92.12 | 92.57 |

As shown in Table 8, as $\bar{C}$ gets smaller, the overall performance increases. The best overall result was obtained for $\bar{C} = 1$ except the sensitivity result, which was marginally less than $\bar{C} = C_{in}/8$ by 0.72%. Over-parameterization (i.e., in this case too many intermediate channels) leads to overfitting on the training set and a subsequent performance decline on the testing set. Additionally, the LIDC-IDRI dataset used in this study is much smaller than the generic datasets used by [18,38], which could further worsen the effects of the overfitting.

## 5.3. The influence of the number of base and extended CAN blocks

In this section, we first conduct an ablation study to examine the optimal number of base feature extractors or base CAN blocks, $f_0$ (see Section 3.3) to use in our final network architecture. We also perform a second (similar) ablation study to analyze the optimal number of extended CAN blocks, $f_i$ added through our new Progressive Growing method proposed in Section 3.3.

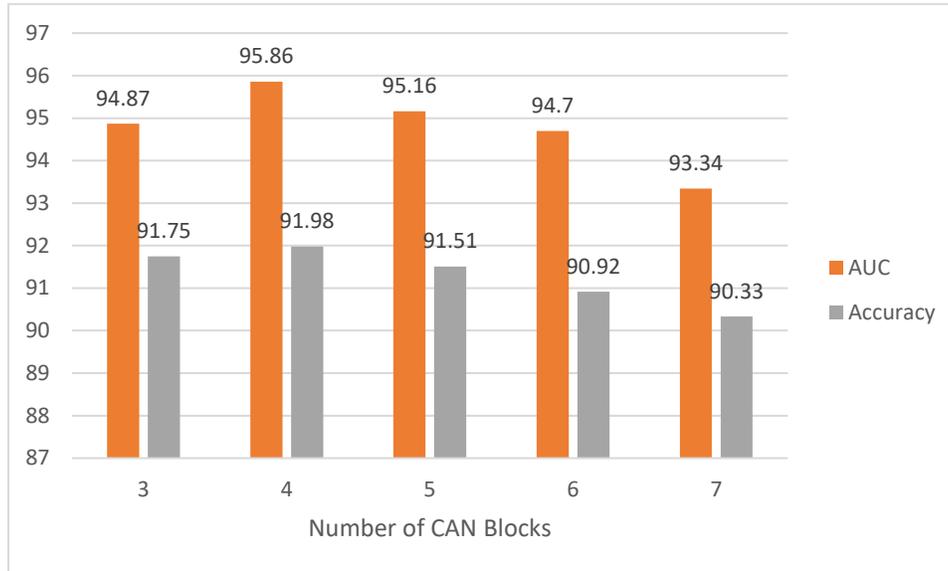

*Figure 3: Effects of varying the number of base CAN blocks, $f_0$ in the network architecture*

First, to find the optimal number of base CAN blocks, $f_0$, we tested our CAN network without using the Progressive Growing algorithm and without Curriculum Learning. The obtained AUC and accuracy results are plotted in Figure 3. We can observe that the accuracy reaches the peak performance when we use four base CAN blocks and the performance degrades as we increase the number of blocks beyond four.

Next, we used the four base CAN blocks as the base feature extractor, $f_0$ and tried a different number of extended blocks, $f_i$ using our Progressive Growing and training method described in Algorithm 1 (see Section 3.4). We examined the results of adding one to four CAN blocks on the fly using the training procedure described in Algorithm 1. The AUC and accuracy results of our ablation study are plotted in Figure 4. From Figure 4, we observe that the best results were obtained using 3 extended blocks.

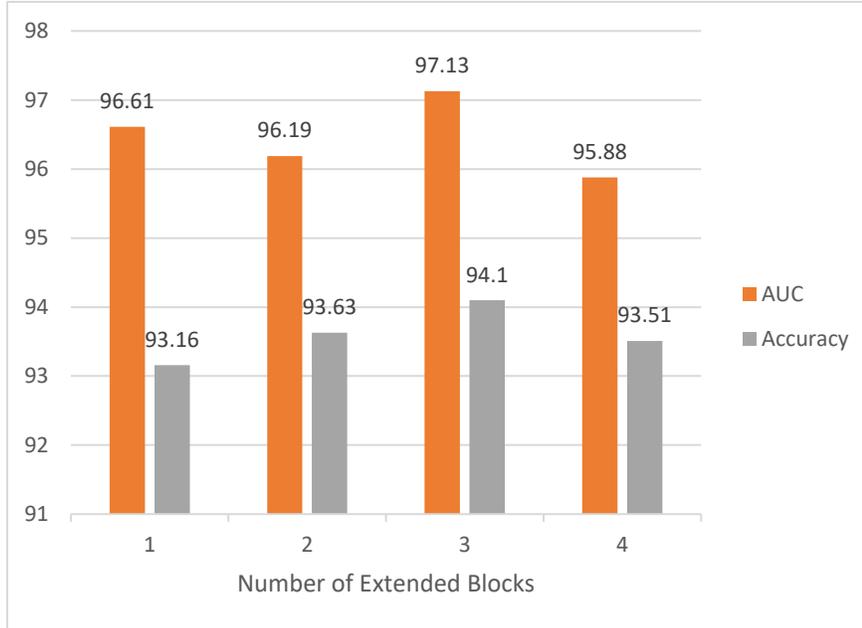

*Figure 4: Effects of varying the number of the extended CAN blocks, $f_i$ in the network architecture*

We also observe that using our Progressive Growing method to extend the CAN blocks to our ProCAN network architecture is very effective as all the (AUC and accuracy) results in Figure 4 outperform the best result of AUC = 95.86% and accuracy = 91.98% obtained for the number of base CAN blocks ($f_0 = 4$) in Figure 3. This result demonstrates that our new Progressive Growing method using the **Bernoulli** matrix instead of just the scalar $p$ effectively enables our network architecture to progressively grow it in a "safe" environment (see also Progressive Growing Ablation study results in Section 5.5 below).

## 5.4. Effectiveness of curriculum learning method

We also conducted an ablation study to analyze the effects of applying different curriculum learning criteria as explained in Section 3.2. The first criterion was based on the malignancy rating of the radiologists and the second criterion on the nodule diameter. We tabulate the results of the ablation study in Table 9.

The ablation study results show that incorporating both the radiologists' ratings and nodule diameter criteria improves the performance over not using Curriculum learning at all. The experiments also show that curriculum learning using the diameter criteria is better than using radiologists' ratings on all performance metrics except the precision. This indicates that the nodule diameter criterion carries more relevant information than the radiologists' ratings for the classification task, which was also observed in our previous study [8].

*Table 9: Effects of applying different curriculum learning criteria on ProCAN's performance. None indicates that no curriculum learning was applied.*

| Criteria | AUC | Accuracy | Precision | Sensitivity | F1-Score |
|---|---|---|---|---|---|
| None | 95.32 | 92.92 | 92.61 | 92.61 | 92.61 |
| Rating | 96.54 | 93.63 | **94.67** | 91.87 | 93.25 |
| Diameter | **97.13** | **94.11** | 94.54 | **93.12** | **93.81** |

*5.5. Effectiveness of progressive growing method*

In the last ablation study, we conducted experiments to show the effect of different blending methods while growing the network as described in Section 3.3 on Progressive Growing. Specifically, we analyzed the results with no blending strategy as in equation (15). We also examined the performances of blended learning using the scalar value, $p_i$ as in the ProGAN method [14] and the newly proposed 2D ***Bernoulli*** matrix (see Section 3.3). We tabulated all the results in Table 10.

*Table 10: Effects of different blending methods on ProCAN's performance. None means no blending was applied as in equation (15)*

| Blending | AUC | Accuracy | Precision | Sensitivity | F1-Score |
|---|---|---|---|---|---|
| None | 94.22 | 89.98 | 90.84 | 87.93 | 89.36 |
| Scalar value, $p_i$ | 95.93 | 93.16 | 92.23 | **93.6** | 92.91 |
| 2D ***Bernoulli*** matrix | **97.13** | **94.11** | **94.54** | 93.12 | **93.81** |

We observe from Table 10 that not applying any blending strategy results in performance dropping significantly/ drastically. As we discussed in Section 3.3, suddenly adding a new layer during the training procedure significantly affects the overall network performance. The results also show that using the new 2D ***Bernoulli*** matrix instead of just the scalar value, $p_i$ improves the results significantly.

*5.6. Effectiveness of batch size*

Finally, we conducted experiments to show the effect of different batch sizes on ProCAN's performance. In Table 11, we observe that 256 is the best batch size across all performance metrics excluding sensitivity, followed by 128 and 64. Our proposed method uses Batch Normalization [27], which is very sensitive to batch sizes. Usually, Batch Normalization works better with larger batch sizes as when the batch size increases, the sample mean and standard deviation in Batch Normalization approximates that of the population more closely [27].

*Table 11: Effects of different batch sizes on ProCAN's performance*

| Batch size | AUC | Accuracy | Precision | Sensitivity | F1-Score |
|---|---|---|---|---|---|
| 64 | 95.95 | 93.75 | 93.37 | 93.60 | 93.48 |
| 128 | 96.45 | 93.99 | 93.40 | **94.09** | 93.74 |
| 256 | **97.13** | **94.11** | **94.54** | 93.12 | **93.81** |

## Conclusion

In this study, we proposed a new Progressive Growing Channel Attentive Non-Local Network (ProCAN) network for lung nodule classification with several novel characteristics. First, we added a new channel-wise attention capability to the highly popular Non-Local network called the CAN block. The CAN block gives our ProCAN network the ability to detect size-invariant nodules with both spatial and channel-wise attention. Second, we developed a

Curriculum Learning method based on the nodule diameter and radiologists' ratings, and found that we can improve model performance if we train it on easy examples (i.e., small or big nodules in the nodule size category and clearly benign or clearly malignant in the radiologist ratings' category) before the difficult examples (i.e., mid-sized nodules in the nodule size category and probably benign or probably malignant in the radiologist ratings' category). Third, we found that progressively growing the network during the training is crucial and improves the results compared with using fixed number of blocks in the overall network architecture. Moreover, we propose a new strategy to blend the new blocks gradually with a matrix sampled from a 2D *Bernoulli* distribution and show that this improves the overall accuracy and the AUC. We analyzed the effectiveness of all these new components/contributions in extensive ablation studies and in comparisons with other state-of the-art methods. The overall experimental results show that our proposed ProCAN model achieves state-of-the-art results in the literature. We will use ProCAN principles to design an end-to-end early-stage lung cancer detection and classification model in future work, and examine its performance on bigger, more diverse datasets.

## Acknowledgements

This work was supported by the Fundamental Research Grant Scheme (FRGS), Ministry of Education Malaysia (MOE), under grant FRGS/1/2018/ICT02/MUSM/03/1. This work was also supported by the Electrical and Computer Systems Engineering and Advanced Engineering Platform, School of Engineering, Monash University Malaysia. This work was also supported by the TWAS-COMSTECH Joint Research Grant, UNESCO.